\documentclass{ws-ijmpa}
\usepackage[super,compress]{cite}
\usepackage{graphicx}
\usepackage{color}
\begin{document}
\markboth{Mathews, Boccioli, Hidaka, and Kajino}{Relic Neutrinos and the Equation of State}

%
\catchline{}{}{}{}{}
%
\title{Review of Uncertainties in the Relic Neutrino Background and Effects from the Nuclear Equation of State}

\author{G. J. Mathews and L. Boccioli
}

\address{Center for Astrophysics, Department of Physics, University of Notre Dame, Notre Dame, IN 46556, USA\\
gmathews@nd.edu, lbocciol@nd.edu}

\author{J. Hidaka}
\address{Mechanical Engineering Department, Meisei University,
   Tokyo 191-8506, Japan\\
National Astronomical Observatory of Japan  Tokyo, 181-8588, Japan\\
jun.hidaka@meisei-u.ac.jp}

\author{T. Kajino }
\address{National Astronomical Observatory of Japan Tokyo, 181-8588, Japan\\
International Research Center for Big-Bang Cosmology and Element Genesis, and School of Physics, Beihang University, Beijing 100083, Peoples Republic of China\\
Graduate School of Science, The University of Tokyo,  Tokyo, 113-0033, Japan\\
kajino@buaa.edu.cn, kajino@nao.ac.jp}
\maketitle

\begin{history}
\received{Day Month Year}
\revised{Day Month Year}
\end{history}

\begin{abstract}
We review the computation of and associated uncertainties in the current understanding of the  relic neutrino background due to  core-collapse supernovae, black hole formation and neutron-star merger events.  We consider the current status of uncertainties due to  the nuclear equation of state (EoS), the progenitor masses, the  source supernova neutrino spectrum, the cosmological star formation rate,   the stellar initial mass function, neutrino oscillations, and neutrino self-interactions.  We summarize the  current viability of future neutrino detectors to distinguish the nuclear EoS and the temperature of supernova neutrinos via the detected relic supernova neutrino spectrum.
\keywords{diffuse radiation - neutrinos - stars: formation - stars: massive - supernovae: general - stars: supergiants.}
\end{abstract}

\ccode{PACS numbers:26.50.+x, 97.60.Bw, 97.60.Jd, 26.60.Kp}


\section{Introduction}\label{sec:Intro}
For a number of   years there has been interest\cite{Hartmann97, Totani96, Beacom04,Beacom06,Beacom10,Yuksel07,Horiuchi09,Lunardini03,Lunardini09,Lunardini10,Yang11,Lunardini12,Keehn12,Chakraborty11,Nakazato13,Mathews14,Hidaka16,Hidaka18,Nakazato15,Mirizzi15,Wei16,Barranco18,Anandagoda20} in the possibility of detecting the cosmic background due  to supernova  relic neutrinos (SRN).  This is also referred to as the diffuse supernova neutrino background\cite{Beacom10} (DSNB).  Although there are other contributions to the diffuse neutrino background from neutron star merger events (NSMs) associated with neutron-star + neutron-star binaries or black-hole + neutron star binaries,\cite{Kyutoku17,Schilbach18,Lin20} the dominant contribution to the diffuse neutrino background is from core collapse supernovae.\cite{Mathews14}

Massive stars ($M \ge 8$ M$_{\odot}$) culminate their evolution as either a core collapse supernovae (CCSNe) or as a failed supernova (fSNe) leading to a 
black-hole remnant.    In either case, such explosive events are expected to produce 
 an intense flux of neutrinos with total energy of order $\sim 10^{53}$ ergs emitted over an interval of $\sim 10$ s for CCSNe or $\sim 1$ s for fSNe.  

Neutrinos are weakly interacting elementary particles, and therefore can emerge 
from  within the interior of CCSNe, fSNe, and NSMs.  As such,  neutrinos 
can provide  information regarding  the physical processes that take place inside these explosive environments.
The detection of this diffuse relic neutrino background and the associated  energy spectra could provide information on
neutrino properties such as flavor oscillations \cite{Chakraborty11,Lunardini12} and/or the neutrino temperatures  produced in CCSNe explosions.\cite{Mathews14} One may even be able to discern \cite{Nakazato13} the shock revival time from the detected spectrum of relic neutrinos.
 Moreover,  these neutrino emission processes have continued over the history of the Galaxy.  Hence,  a detection of the SRN background will not only probe the physics of CCSNe, fSNe, and NSMs, but also the history of explosive events in the Galaxy,\cite{Hopkins06} as well as the cosmic expansion history itself.\cite{Barranco18}. The possibile interaction of neutrinos with the diffuse supernova neutrino background also constrains beyond standard-model physics.\cite{Shashank19}

The prospect for detecting the SRN background in the near future seems possible due to plans\cite{abe11,Takeuchi20} for a next generation Hyper-Kamiokande detector with a mega-ton of pure water laden with 0.1-0.2$\%$ GdCl$_3$ to enhance the neutron tagging efficiency and reduce the background.\cite{Beacom04}  Moreover, results from the Super-Kamiokande collaboration \cite{SK12} already place some marginal limits on the total neutrino energies and temperatures from the background spectrum of SRN.\cite{SK12,Sek13} The possibility of a liquid scintillation detector has also been discussed.\cite{Wei16}

However, as analyzed in a number of works (e.g. Refs.~[\refcite{Mathews14,Hidaka16,Hidaka18,Nakazato15}]), theoretical predictions of the detection rate of supernova relic neutrinos are still subject to 
a number of  uncertainties. 
These uncertainties include   the time dependence of the stellar initial mass function, the star formation rate and/or  supernova rate, the roles of neutrino oscillations and/or neutrino self interaction, and 
the spectral energy distributions of the three flavors of supernova neutrinos.  {These  in turn  depend upon the explosion mechanism,  the equation of state (EoS) for proto-neutron stars, and whether the final remnant is a neutron star or black hole.
 In this review we summarize the current status of some of these uncertainties. In particular we highlight the new trends toward decreasing uncertainty due to the equation of state and  the supernova neutrino spectrum and luminosity.}

\section{The Relic Neutrino Background}

 The presently arriving  cosmic diffuse neutrino flux spectrum  ${dN_\nu}/{dE_\nu} $ can be derived \cite{Hartmann97,str05,Yuksel08} from an integral over the  cosmic
 redshift $z$ of the neutrinos emitted per event (SN, or NSM)  and the cosmic event rate  per redshift per comoving volume ($R_{SN}(z)$ and $R_{NSM}(z))$:
\begin{eqnarray}
 \frac{dN_\nu}{dE_\nu} &=&
 \frac{c}{H_0}\int_0^{z_{max}}\biggl[R_{SN}(z)\frac{dN^{SN}_\nu(E^\prime_\nu)}{dE^\prime_{\nu} }+ R_{NSM}(z)\frac{dN^{NSM}_\nu(E^\prime_{\nu})}{dE^\prime_\nu}\biggr] \nonumber \\
 &\times&
 \frac{dz}{\sqrt{\Omega_m(1 + z)^3 + \Omega_\Lambda}} \label{4-15}~~,
\label{nuflux}
\end{eqnarray}
where $z_{max} \approx 5 $ is the redshift at which star formation begins.  The various contributions to $R_{SN}$ are described below, while $R_{NSM}$ is the corresponding number of NSM events.  The quantities  $dN^{SN}_\nu(E^\prime_\nu)/dE^\prime_\nu$ and $dN^{NSM}_\nu(E^\prime_\nu)/dE^\prime_\nu$ are the emitted neutrino spectra at the various sources , where the energy  $E'_\nu= (1+z)E_\nu$ is the energy at emission, while $E_\nu$ is the redshifted energy to be observed in the detector.   
In addition to integrating over the  range of progenitor models to determine $R_{SN}$ for normal CCSNe, one should add\cite{Mathews14} the neutrino emission from ONeMg electron-capture  supernovae, the  neutrino emission from the collapse of massive stars that form black holes, and the contribution from NSMs.  The contribution from NSMs is small but potentially interesting.  The reader is referred to Refs.~[\refcite{Kyutoku17,Schilbach18}] for discussions of this possibility.  For the remainder of this review we will 
only consider contributions from various classes of supernovae or black-hole formation.

\begin{figure}
\centerline{\includegraphics[width=3.5in]{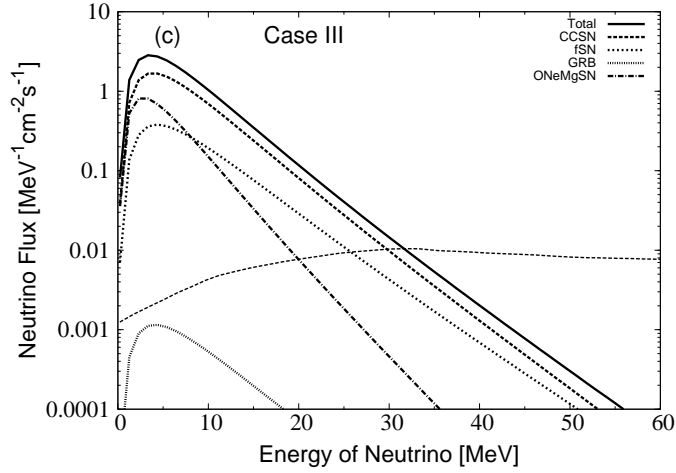}}
\caption{Illustration\cite{Mathews14} of the relative contributions to the total 
 predicted  SRN flux spectrum (Solid line) from different sources in the case of no neutrino oscillations (Case III).  These are:  CCSNe (short dashed line);  ONeMg SNe (dot-dashed line); fSNe (dotted line);  and Collapsar GRBs (small dotted line).   The background due to atmospheric neutrinos is also indicated as a short dashed line. 
\label{fig:1}}
\end{figure}

Figure \ref{fig:1} from Ref.~[\refcite{Mathews14}] illustrates the arriving neutrino flux from different supernova sources as labelled.   This figure is for the case of no neutrino oscillations (Case {\it III} as defined below).  We next review the various factors in Eq.~(\ref{nuflux}) in detail.

\subsection{Core-Collapse Supernova Rate $R_{SN}$}
The core-collapse supernova rate $R_{SN}$ can be determined from the total star formation rate (SFR) [i.e. total mass in stars formed per year per comoving volume] if one  determines the fraction of total stars that produce each observable class of supernovae or other sources.  The general form for the supernova rate at a given redshift $z$ is related to the star formation rate $\psi_*(z)$ and the initial mass function (IMF) $\phi(M)$.
\begin{equation}
 R_{SN}(z) =  \psi_*(z) \times \frac{\int_{min}^{max}dM \phi(M)}{\int_{M_{min}}^{M_{max}}dM
   M\phi(M)} ~~,
\label{RSN}
\end{equation}
where $min$ and $max$ denote the range of progenitors for supernovae.  As described below this may be a sum of segments that lead to CCSNe or fSNE separately.  The quantities  $M_{min}$ and $M_{max}$ refer to the range over all stars formed.

It is often assumed\cite{Horiuchi11} that all massive progenitors  from 8 to 40 M$_\odot$ lead to visible SNe.  The higher end of this mass range corresponds to SNIb,c, while the lower end corresponds to normal SNII.   

In Ref.~[\refcite{Mathews14}], however, it was pointed out  contributions from both normal core-collapse SNII and  SNIb,c should be   considered separately, along with the possibility  of fSNe forming a black hole and dim {ONeMg} supernovae in the low-mass end.  The SNIb,c  events accounts for about $\sim 25 \pm 10$\% of observed core-collapse supernova rate at redshift near $z=0$ \cite{Smartt09}. 
Hence, for the total observed core-collapse  SN rate one can write:
\begin{equation}
 R_{SN}(z) = R_{SNII}(z) + R_{SNIb,c}(z)
\label{RSNtot}
\end{equation}
 Nevertheless, there is large uncertainty in the mass ranges of CCSNe and the degree to which progenitors have experienced mass loss such that the true initial progenitor mass could be greater. 

 Observational evidence for the mass of SN progenitors has been obtained \cite{Smartt09a,Smartt09} by comparing sky images of nearby SNe before and after explosion. This suggests that the maximum mass of the red supergiant progenitors of  core-collapse SNe may be as small as 16.5 $M_{\odot}$. This is the so-called  "red supergiant(RSG) problem''\cite{Smartt09a,Smartt09}. A subsequent study,\cite{Smartt15} however, suggested that this value could be as much as 18 $M_{\odot}$.  A similar range was deduced in Ref.~[\refcite{Sukhbold19}] who argued that this limit is closely related to the transition from convective to radiative transfer during the central carbon-burning phase.  Hence, 16.5 to 18 $M_\odot$ was adopted in Ref.~[\refcite{Hidaka16}] as a reasonable range of the uncertainty in the lower mass limit for fSNe.
 
 It is also believed  \cite{Heger03} that initial progenitor stars in the mass range of 8-10 M$_\odot$ collapse as an electron degenerate ONeMg core and do not produce bright supernovae.  Hence,  one should consider a lower limit to the progenitor of normal CCSNe of 10 M$_\odot$ when allowing for the possibility of ONeMg SNe.  

Regarding SNIb,c, although it is known \cite{Smartt09} that they are associated with massive star forming regions, there is some ambiguity as to their progenitors.  There are two theoretical possibilities. Refs.~[\refcite{Heger03,Fryer99}] have studied the possibility that massive stars ($M \sim 25-100$ M$_\odot$) can  shed their outer envelope by radiative driven winds leading to bright SNIb,c explosions.  This source for SNIb,c was adopted in \cite{Horiuchi11} by assuming that all stars with  ($M \sim (25-40) $ M$_\odot$) explode as SNIb,c, while   Ref.~[\refcite{Hopkins06}] adopted a range of (25-50) M$_\odot$ for SNIb,c.  Changing the upper mass limit from 40 M$_\odot$ to 50 M$_\odot$, however,  makes little difference (~1\%) in the inferred total core-collapse supernova rate.\cite{Mathews14} 

Theoretically,  however, this single massive-star paradigm does not occur as visible SNIb,c until well above solar metallicity.\cite{Heger03}  At lower metallicity  most massive stars with $M > 25$ M$_\odot$ collapse as failed supernovae with black hole remnants.  Hence, this mechanism is not likely to contribute to the observed SNIb,c rate at higher redshifts.  On the other hand, it has been estimated \cite{Fryer99} that 15 to 30\% of massive stars up to $\sim 40$ M$_\odot$ are members of interacting binaries that can shed their outer envelopes by Roche-lobe overflow \cite{Podsiadlowski92, Nomoto95} also leading to bright SNIb,c events.  That hypothesis was considered in Ref.~[\refcite{Mathews14}].  That is, a  fraction $f_b$ of all massive stars in the range of (8 to 40) M$_\odot$ could  result in bright SNIb,c via binary interaction.  
 
\subsection{Failed Supernovae}
It is usually presumed\cite{Heger03} that  all single massive stars with progenitor masses
with $M > 25$ M$_{\odot}$ end their lives as failed supernovae fSNe.  However,  this cutoff is uncertain.  For single stars above this cutoff mass   the whole star collapses leading to a black hole and no detectable supernovae. 
Allowing for a fraction $f_b \sim$ 25\% of stars with $M  = (25-40)$ M$_\odot$ to form SNIb,c by binary interaction, then 
the rate for failed supernovae is given by,\cite{Mathews14}
\begin{eqnarray}
 R({\rm fSN})  &= & (1-f_b) \psi_*(z) \times \frac{\int_{25 M_\odot}^{40M_\odot}dM \phi(M)}{\int_{M_{min}}^{M_{max}}dM
   M\phi(M)} ~~ \nonumber \\
   &&+  \psi_*(z) \times \frac{\int_{40M_\odot}^{125 M_\odot}dM \phi(M)}{\int_{M_{min}}^{M_{max}}dM
   M\phi(M)} ~~.
 \label{4-5-3}
\end{eqnarray}

There appears to be a non-monotonic  dependence of the success  of the explosion on  progenitor mass  \cite{Ugliano12,Pejcha15,Sukhbold16,Couch18}. These numerical investigations suggest  that the  compactness of the progenitor core just prior to collapse is a good indicator of the success of the explosion, although the onset of supernova turbulence may be a better indicator.\cite{Couch18}.  A correct estimate of the relic neutrino background should include segmented ranges for failed supernovae and successful supernovae in the range from 10 to 25 M$_\odot$.

\subsection{ONeMg Supernovae}
ONeMg (or electron-capture) supernovae involve  explosion energies and luminosities that are an order of magnitude less than normal core collapse supernovae.  Hence, they  may go undetected. ONeMg supernovae are believed to arise from progenitor masses  in the range
of 8 $M_{\odot} \le M \le 10 M_{\odot}$.  Such supernovae occur \cite{Isern91} when an electron-degenerate ONeMg  core reaches a critical density 
(at $M_{core} \sim 1.4$ M$_\odot$).  In this case the electron Fermi energy exceeds the threshold for electron  capture on $^{24}$Mg.  These electron captures  cause a loss of hydrostatic support from the electrons and also a heating of the material.   Ultimately, this leads to a dynamical collapse and the ignition of an O-Ne-Mg burning front that consumes the star.  

 \subsection{Star formation rate}
 \label{SFR}

The cosmic star formation rate $\psi_*(z)$ is key component of theoretical estimates of the  SRN background. Determining the SFR is an involved procedure\cite{Horiuchi11,Mathews14,Hidaka18} based upon different sources, e.g., UV light from galaxies and the far infrared (FIR) luminosity \cite{Madau14}.  
It also depends {\cite{Conroy13}} upon modeling  fits to the spectral energy distribution (SED). Uncertainties from this procedure were analyzed  in Refs.~[\refcite{Mathews14,Hidaka18}].
 The star formation rate also depends upon the choice of the stellar initial mass function (IMF). 

 It was noted in Ref.~[\refcite{Horiuchi11}] 
that the measured core-collapse supernova rate [$R_{SN}({\rm Obs})$]  in the redshift range 0$\le z \le$1 
appears to be about a factor of two smaller than the core-collapse supernova rate deduced from the measured\cite{Madau14} cosmic massive-star formation rate (SFR).  This is the so-called supernova rate problem\cite{Mathews14,Horiuchi11}. One possible solution to this involves uncertainty\cite{Mathews14} in the inferred supernova rate out to redshift $z \sim 1$.  Another possible explanation is from the uncertainty in the lower mass limit for the progenitors of CCSNe.\cite{Hidaka16}

A commonly adopted form for a fit to the SFR is the  Madau formula \cite{Madau14} given by:
\begin{equation}
\label{eqn:SFR_madau}
\psi_{*}(z)=\frac{a (1+z)^{b}}{1+[(1+z)/d]^{c}}\, \rm{M_{\odot}} \rm{year}^{-1}\rm{Mpc}^{-3}, 
\end{equation}
where $a= 0.015$, $b= 2.7$, $c= 5.6$, and $d= 2.9$.

\begin{figure}[h]
\begin{center}
\includegraphics[width=3.in]{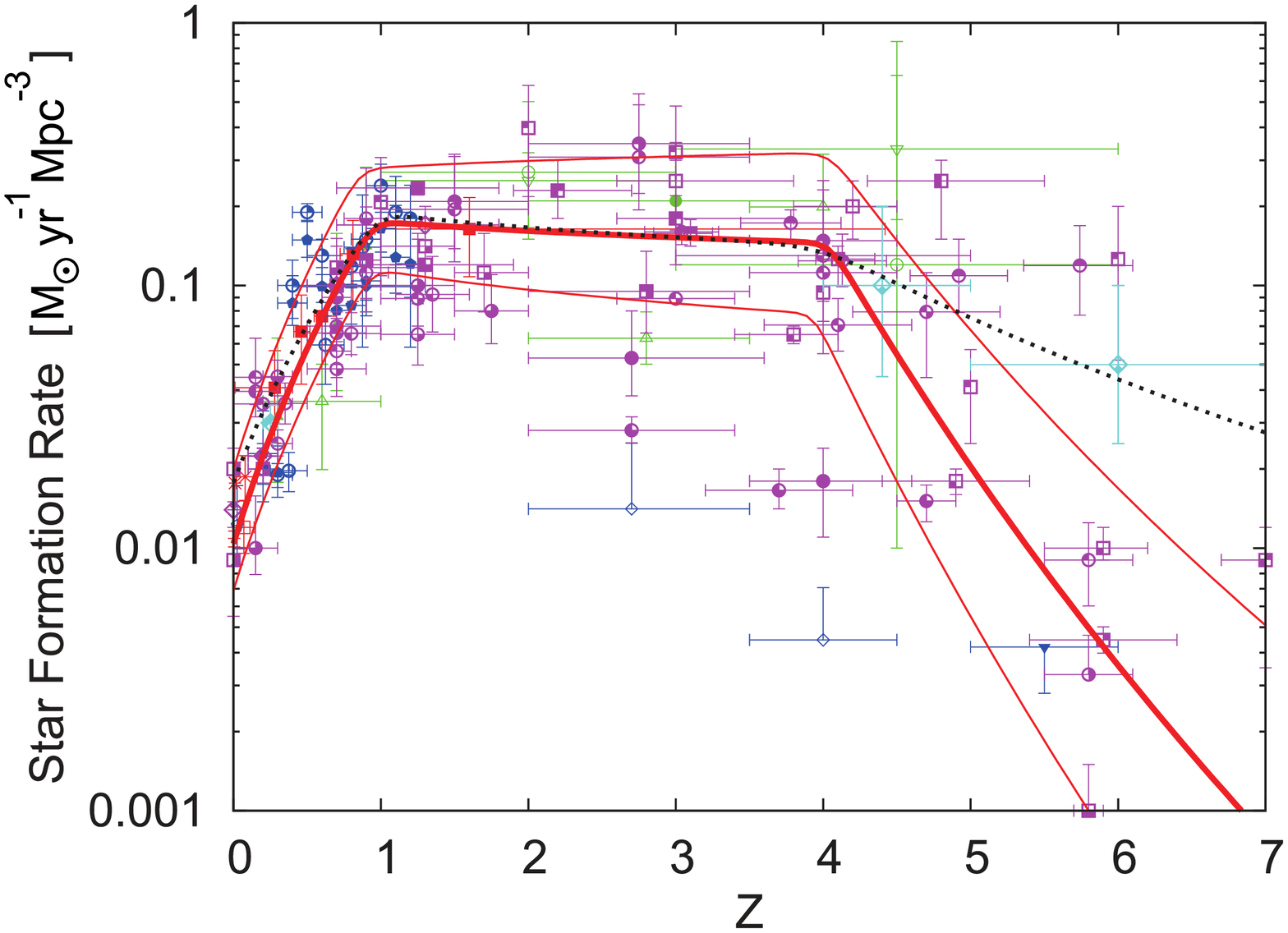}
\includegraphics[width=3.in]{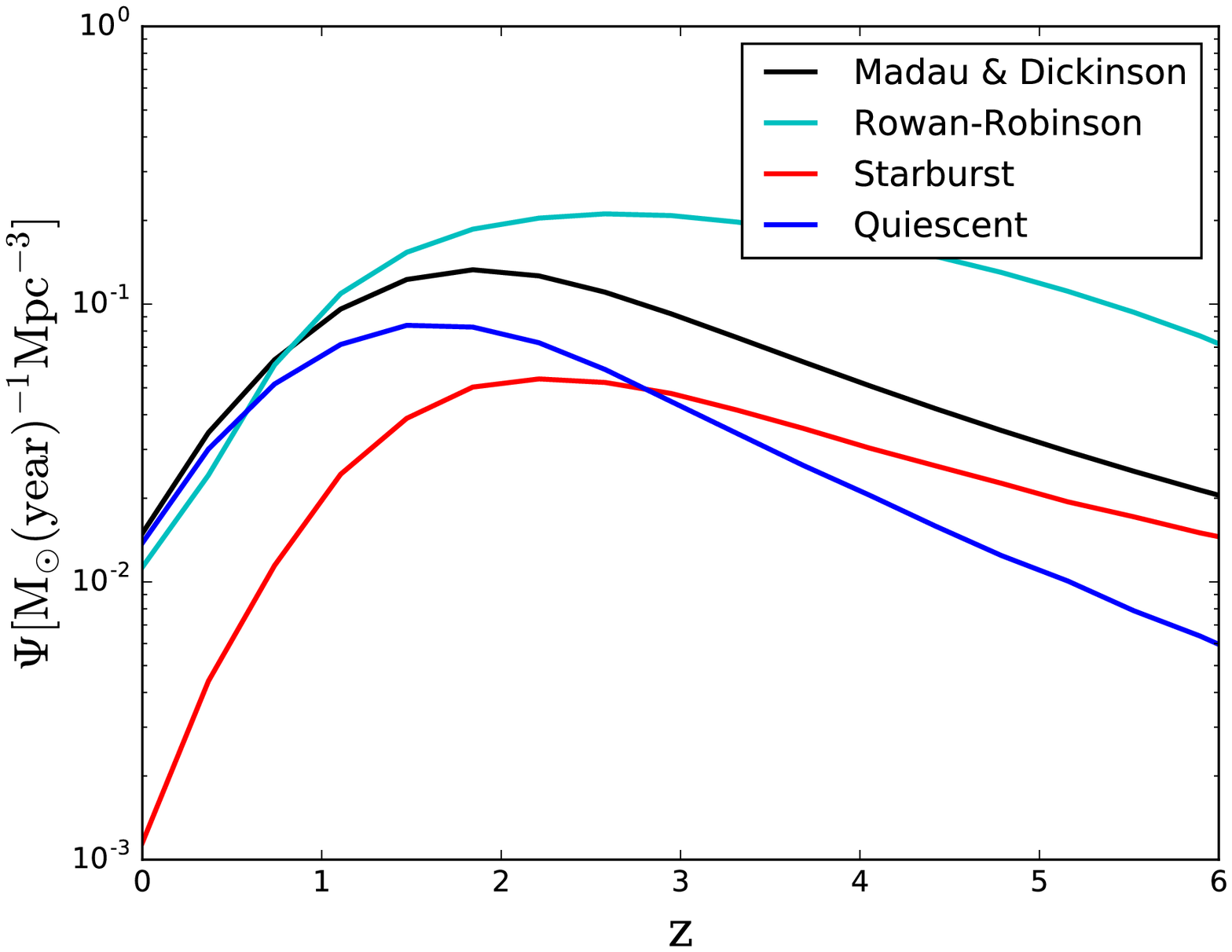}
\caption{The top panel shows a piecewise linear  star formation rate function from a fit\cite{Mathews14} to the subset of  observed dust-corrected data.
Red, blue, magenta, light blue, and green points show the observed data in IR,
optical, UV,  X-ray / $\gamma$-ray, and  radio bands, respectively. 
 Red lines show the SFR as  function of redshift $z$ deduced from $\chi^2$
fitting, along with the $ \pm 1\sigma$ upper and lower limits to the  SFR.  The reduced $\chi^2_r$ for the fit is 2.3. 
The black dotted line shows the star formation rate at high redshift proposed in Y\"uksel et al.\cite{Yuksel08}   The bottom panel from Hidaka et al.\cite{Hidaka18} illustrates the uncertainty in the star formation rate from various parameterizations.  The black line is the  commonly adopted Madau SFR.\cite{Madau14}. The cyan line shows the SFR of\cite{RR_SFR_2016}. For illustration the  figure also includes possible additional starburst (red) and quiescent (blue) components as suggested by Lacey et al.\cite{Lacey16}.
}
\label{fig:2}
\end{center}
\end{figure}

Another convenient  functional form for the SFR is conveniently parameterized as a continuous piecewise linear  fit to the observed star formation rate vs. redshift.\cite{Yuksel08} 
\begin{equation}
\psi_{*}(z)=\dot{\rho_0}[
(1+z)^{\alpha \eta}+(\frac{1+z}{B})^{\beta \eta}+(\frac{1+z}{C})^{\gamma \eta}
]^{\frac{1}{\eta}}.
\label{eqn:psi_*}
\end{equation}
Parameters adopted in Ref.~[\refcite{Yuksel08,Horiuchi11}] are  $\dot{\rho_0} = 0.016$ $h^3_{73}$  M$_\odot$ Mpc${-3}$ yr$^{-1}$ for the cosmic SFR at $z = 0$, as well as the parametrization $\alpha = 3.4$, $\beta=-0.3$, $\gamma=-3.5$, $z_1 =1$, $z_2 = 4 $ are smoothed with $\eta=-10$, and the normalization is given by $B = (1-z_1)^{\alpha/\beta}=5000$ and $C = (1+z_1)^{(\beta-\alpha)/\gamma}(1+z_2)^{1-\beta/\gamma}=9$. A somewhat different parameterization was deduced in [\refcite{Mathews14}] based upon different data at low redshift. The piecewise fit from [\refcite{Mathews14}]
 is illustrated in the upper panel of Figure \ref{fig:2}.  The fit from [\refcite{Yuksel08}] is shown by the dotted line.  
 
 As an illustration of the overall uncertainty in the derived star formation rate the bottom panel  of Figure \ref{fig:2} (from Hidaka et al.\cite{Hidaka18}) shows  the star formation rate from various parameterizations.  The black line is the  commonly adopted Madau parameterization\cite{Madau14} of Eq. (\ref{eqn:SFR_madau}). The cyan line shows the SFR of Rowan and Robinson\cite{RR_SFR_2016}. For illustration the  figure also includes possible additional starburst (red) and quiescent (blue) components as suggested by Lacey et al.\cite{Lacey16}.  The effects of these additions to the SFR were explored in Ref.~[\refcite{Hidaka18}] and found to be not particularly important for determining the SRN spectrum.   The reason for this\cite{Mathews14}  is because the observed relic supernova neutrino spectrum is  mainly determined by the SFR for $z \le 1$.  Although the the star formation rates  at high redshift can deviate by more than an order of magnitude, the overall uncertainty in the relic supernova neutrino spectrum due to the SFR remains less than a factor of 2.\cite{Mathews14}

Most evaluations of the  SFR are based upon the UV luminosity that is mainly due to OB stars. 
The dust surrounding galaxies, however, always complicates the evaluation of the UV luminosity. There is a claim that the SFR based upon UV light is underestimated especially for  starburst galaxies \cite{RR_SFR_2016}.  A method based upon  SED modeling has been proposed\cite{RR_SFR_2016}  that takes into account the star formation embedded in dense molecular clouds. This study indicated that the SFR for  $z>3.5$ is higher by a factor of 2 to 3 than the estimate from  UV light.  Effects of this were analyzed in  [\refcite{Hidaka18}].

In Ref.~[\refcite{Mathews14}]  three different versions of the star formation rates were studied as a means to estimate the overall uncertainty in the SFR dependence.  One was
a revised  SFR
 based upon a piecewise linear  fit to the observed star formation rate vs. redshift. They also considered SFR models\cite{kob00} with and without corrections for  dust extinction.
 
\subsection{Initial mass function $\phi(M)$}

 A   convincing theory for the universal IMF has not yet been established. One of the remaining uncertainties is the metallicity dependence of the IMF.  In Ref.~[\refcite{Hidaka18}]  the dependence of the detected neutrino spectrum  on a metallicity dependent IMF and other changes in the IMF were analyzed. 
A broken power-law\cite{Baldry03} Salpeter A IMF (Sal-A) is often assumed,
\begin{equation}
\phi(M) = M^{-\zeta} ~~,
\end{equation}
with $\zeta = 2.35$ for stars with $M \ge 0.5$ M$_\odot$ and $\zeta = 1.5$ for $0.1$ M$_\odot  < M < 0.5$ M$_\odot$.  

 Most of the popular IMFs including the Sal-A are observationally deduced for stars in the Milky Way. There are also theoretical derivations of the power-law index for the IMF. For example,  a power law IMF is explained by the turbulent motion of molecular clouds in a magnetic field.\cite{Padoan02} 

The relationship between metallicity and the IMF  has also been considered in [\refcite{Hidaka18}]. The IMF in metal poor environments could favor higher mass stars  because  molecular clouds  may not sufficiently cool  to allow gravitational  collapse  on small scales. Ref.~[\refcite{Hidaka18}] explored  the IMF-metallicity relationship by adopting a cosmological metallicity evolution. This was based upon  observational evidence\cite{Martin-Navarro_2016} that  IMF evolution occurs in low-$z$ galaxies for which the  IMF-Metallicity relationship obeys   
\begin{equation}\label{eqn:z-dependent_IMF}
\zeta = 2. 2(\pm 0.1) + 3. 1(\pm0.5)\times [\rm{M/H}]~.
\end{equation}
Metallicity evolution is closely related with galaxy evolution. However, both the stellar and gas phase metallicity must be accurately considered. Ref.~[\refcite{Ma_2016}] studied the galaxy mass-metallicity relation based upon cosmological zoom-in simulations that include gas inflow and outflow to estimate the metallicity evolution accurately. They found $[M/H]=0.4[\log(M_*/M_{\odot})-10]+0.67\exp(-0.50z)-1.04$, where $M_*$ is the galactic stellar mass. For the metallicity of the gas phase, it was found that $[M/H]=0.35[\log(M_*/M_{\odot})-10]+0.93\exp(-0.43z)-1.05$. In addition, galactic mass evolution is expected, and the average of $M_*$ depends upon the redshift $z$. The galaxy cosmological mass function can be expressed\cite{Lopes_2014} as  $\langle M_* \rangle =10^{11} (1+z)^{-0.58} [M_{\odot}]$. These two studies have been used\cite{Hidaka18}  to deduce the cosmic metallicity evolution and to determine that a $z$-dependent IMF has a small effect on the observed SRN.

\section{Source Neutrino Spectrum $({dN^{SN}_\nu(E^\prime_\nu)}/{dE^\prime_{\nu} })$}
\subsection{Neutrino spectra} 
The uncertainty due to variations of estimates of the emergent neutrino spectra from core collapse supernovae was analyzed in [\refcite{Mathews14}].
 It is usually assumed\cite{Yuksel08,Mathews14} that the dependence on progenitor mass  is small compared to the uncertainty in the neutrino temperatures themselves. 
 The reason for this is  that the mass  of most observed neutron-star supernova remnants  is rather  narrowly
constrained\cite{Lattimer12} to be $\sim 1.4$ M$_\odot$.
This narrow  range of observed remnant neutron star masses
suggests that the associated neutrino temperatures  should   also be tightly constrained. Hence, it is reasonable to adopt  a SN 1987A model (i.e. progenitor mass $\simeq$ 16.2
M$_{\odot}$, remnant mass $\simeq$ 1.4M$_\odot$, liberated binding energy
$\simeq$ 3.0 $\times$ 10$^{53}$ erg  [\refcite{arn89}]).  
Nevertheless, an evaluation of the validity of the assumption that this SN1987A model  is representative of  every core-collapse SN with progenitor masses
in the range of 8 to  25 M$_\odot$ and also every SNIb,c over the  mass range of 8-40 M$_\odot$ in any of the possible paradigms needs to be done.  

The liberated binding energy is divided roughly equally among the  6 neutrino species (3 flavors and their
anti-particles).  The neutrino spectra can be roughly approximated by a Fermi-Dirac  distribution  with temperatures and chemical potentials determined from fits to the numerically determined neutrino transport. 

One can characterize the emergent neutrino spectra by the average neutrino energies (or temperatures and chemical potential) in time.  Even in the context of a single progenitor model, however, it was noted in Ref.~[\refcite{Mathews14}] that  a range of predicted neutrino temperatures and chemical potentials are obtained from  various supernova core-collapse 
simulations in the literature. Neverthe;ess, for most models the fitted chemical potential is small and can be neglected. It was concluded\cite{Mathews14} that the uncertainty ($\sim \pm 50\%$) in the neutrino temperatures constituted one of the largest uncertainties in the expected relic neutrino detection rate.  
However,  a recent survey\cite{OConnor18} of modern supernova collapse simulations all tend to  predict lower uniform neutrino  energies and/or temperatures.   However deviations were noted at late times.

 
 \subsection{Neutrino Luminosities}
\begin{figure}
\centerline{\includegraphics[width=5.5in]{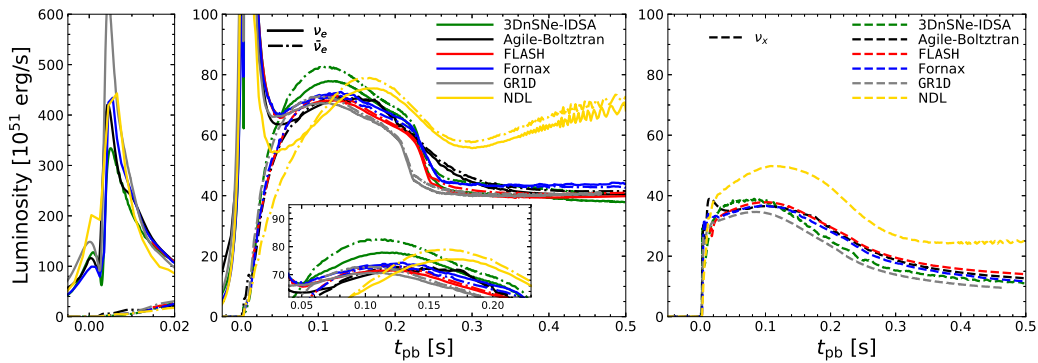}}
\caption{Comparison of neutrino luminosity light curves predicted by various CCSN codes in the literature.\cite{OConnor18}  There is remarkable agreement among most codes except the NDL prediction.  This relates to the use of more extensive\cite{NuLib} neutrino opacities in the newer codes as described in the text.} 
\label{fig:3}
\end{figure}

 The emergent neutrino spectra can be integrated to obtain the total luminosities of each flavor over time.  Figure \ref{fig:3} shows a comparison of a benchmark study with a number of modern spherical 1D SN codes\cite{OConnor18} plus our own results based upon the older general relativistic Wilson (NDL) code.\cite{Wilson03,Olson16}   
 

Figure \ref{fig:3} shows,  that there is now remarkable similarity among the neutrino luminosities for all of the modern codes except for the older CCSN models like the  NDL.\cite{Wilson03,Olson16}  This difference can be traced\cite{Boccioli19} to the use of better  neutrino interaction rates\cite{NuLib} in the newer codes.  Specifically, the NDL code makes use of a simpler formulation of neutrino opacities as described in Refs. [\refcite{Mayle87,Wilson03,Itoh75}], whereas the convergence of the newer CCSN codes shown on Fig. \ref{fig:3} is  based upon a more recent set of neutrino opacities\cite{OConnor18}. Scattering and absorption on free nucleons and heavy nuclei along with electron neutrino absorption on nuclei,  inelastic neutrino-electron scattering and electron-positron annihilation are from Ref. [\refcite{Bruen85}].    Weak magnetism, recoil corrections\cite{Horowitz02} and  ion-ion correlations.\cite{Horowitz97}  and  corrections for the nuclear form factor \cite{Bruenn97,Rampp02} and nucleon-nucleon Bremsstrahlung.\cite{Hannestad98}  In \refcite{Boccioli19} it was shown that in large part, the increase in the opacity due to neutrino-nucleon scattering could account for the difference between the NDL and newer codes. 

Moreover,   in Ref.~[\refcite{Schneider19}] neutrino luminosities were calculated for 97  separately derived equations of state that satisfy nuclear and neutron star constraints.   There is good convergence of the luminosities for all equations of state but the average neutrino energies begin to diverge  after about 1 sec.  Overall, the convergence of the neutrino spectra and luminosities emergent from the more recent neutrino opacities and equations of state suggest that  there is less temperature uncertainty in the SRN spectrum than was  deduced in Ref.~[\refcite{Mathews14}].  

\section{SRN Detection Rate}
A Hyper-Kamiokande next-generation \v{C}erenkov detector has been proposed\cite{abe11,Takeuchi20}, consisting of a mega-ton of pure water laden with 0.1-0.2$\%$ GdCl$_3$ to enhance the neutron tagging efficiency and reduce the background.\cite{Beacom04}
The detected SRN energy spectrum in such a detector can be written  as:
\begin{equation}
\frac{dN_{event}}{dE_{e^+}} = N_{target} \cdot \varepsilon(E_\nu) \cdot
\frac{1}{c} \cdot \frac{dF_\nu}{dE_\nu}
\cdot \sigma(E_\nu) \cdot \frac{dE_\nu}{dE_{e^+}} \label{3-1-1}
\end{equation}
where $N_{target}$ is the number of target particles in the water
\v{C}erenkov detector, $\varepsilon(E_\nu)$ is the efficiency for neutrino
detection, ${dF_\nu}/{dE_\nu}$ is the incident flux of SRNs,
and $\sigma(E_\nu)$ is the cross section for  neutrino absorption:
$\bar{\nu_e} + p \to e^+ + n$, and  $E_\nu = E_{e^+} + \Delta m_{np}$, where $\Delta m_{np} = 1.3~{\rm MeV}$ is the neutron-proton mass difference. For simplicity one can set
$\varepsilon(E_\nu)$ = 1.0, and use the  cross
sections given  in [\refcite{str03b}] when calculating the reaction rate of the SRN with
detector target material.

One only needs to  consider the detection of $\bar{\nu_e}$, because the cross section  in a
water \v{C}erenkov detector for
the $\bar{\nu_e} + p \to e^+ + n$ reaction is about 10$^2$ times larger
than that for $\nu_e$ detection via  $\nu_e + n \to e^- + p$. 
The
 threshold  detection energy
for  SRN  is $\sim10$ MeV due  to the existence of background $\bar{\nu_e}$ emitted
from terrestrial nuclear reactors.    An upper detection  limit is $\sim30$ to 40 MeV due  to the 
existence of noise from the atmospheric $\bar{\nu_e}$ background. 

\begin{figure}
\includegraphics[scale=.50]{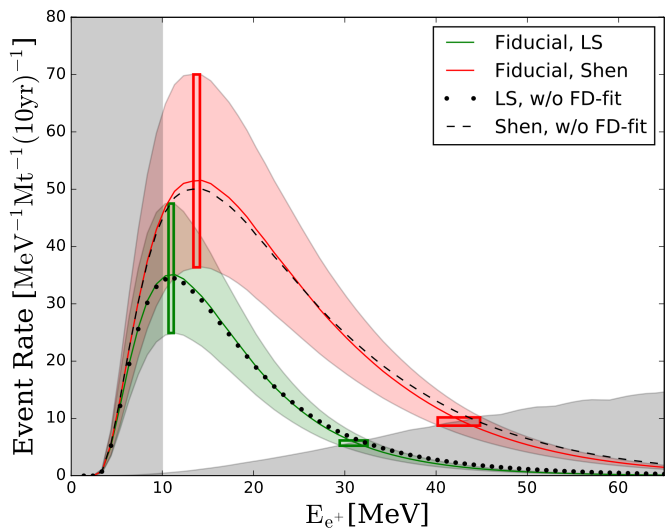}
\caption{Predicted $e^+$ energy spectra and uncertainties\cite{Hidaka16} in the total SRN detections for the case of no neutrino oscillations. The uncertainty bands are based upon the dispersion in the observed cosmic SFR and the detector statistics.  The results for two different fSNe models based upon a stiff Fiducial Shen\cite{Shen98}  EoS (red line)  and a soft Fiducial LS EoS\cite{LS91}  (green line). The dashed and dotted lines illustrate the uncertainty in using a FD fit to the spectra rather than the numerical spectra.   The grey shaded  energy range below 10  MeV indicates the region where  the background noise due to reactor $\bar{\nu_e}$ may dominate. The shaded energy range that intersects the spectrum at   $\sim 30$ to $ 46$ MeV indicates the region where the  background may be dominated by noise from  atmospheric neutrinos. Rectangular regions represent the peak of the spectrum and the locations where the neutrino signals overwhelm the atmospheric neutrino noise. }
\label{fig:4}
\end{figure}

An illustration of the detector SRN spectrum from Ref.~[\refcite{Hidaka18}] is shown in Figure \ref{fig:4}.  This shows the anticipated number of events after 10 years run time.  The vertical grey shaded region below 10 MeV illustrates  the background due to terrestrial reactors.  The lower  grey shaded region indicates the expected background noise due to atmospheric neutrinos.  The red shaded bands indicate the uncertainty in the predicted neutrino signal from the uncertainty in the SFR and the detector statistics.  It was suggested  in Ref.~[\refcite{Hidaka18}] that the largest uncertainty may be due to different models for the EoS for fSNe.  The upper pink band is for a relatively stiff fiducial EoS\cite{Shen98} while the lower green band is the spectrum for a fiducial soft EoS.\cite{LS91}

\subsection{Nuclear EoS Uncertainty}
The results shown in Figure \ref{fig:4} and the EoS uncertainty deduced in Refs.~[\refcite{Mathews14,Hidaka16,Hidaka18}] are based upon the effects of two extreme equations of state.  One is the very soft Thomas-Fermi plus Liquid Drop Model EoS of Lattimer and Swesty\cite{LS91} (for a nuclear compression modulus of $K=180$). 
The other is the rather stiff Relativistic Mean Field (RMF) 1998 model Shen et al.\cite{Shen98}  These were chosen because they represented extreme examples of available calculations  of the neutrino spectra from both CCSNe and fSNe.\cite{Sumiyoshi05,Sumiyoshi08}. 

{It should be noted, however, that there are now more stringent observational limits on the EoS than when those two formulations were developed.  For one, there are now firm determinations\cite{Demorest10,Antoniadis13}  of binary neutron star systems for which a neutron star mass as large a $2.01 \pm 0.04$ M$_\odot$ is obtained.\cite{Antoniadis13}   More recently, the  mass of a neutron star in  the MSP J0740+6620 system was measured\cite{Cromartie19} to be $2.17^{+0.11}_{-0.10}$ M$_\odot$.  Such large neutron star masses can not be achieved with as EoS as soft as that of the Lattimer \& Swesty (K=180) formulation.  

Moreover, another important development is the observation by the LIGO and VIRGO  Scientific collaboration of gravity waves from the binary neutron-star merger GW170817.\cite{Abbot17}. Subsequent analysis\cite{Abbot18} in which both stars are required  to have the same EoS places strong limits on the tidal polarizability which in turn depends upon the neutron star radius to the 5th power.  Indeed, only an EoS for which the radius of a 1.4 M$_\odot$ neutron star is less than 13.3 km is consistent with the 90\% confidence limit.  A separate analysis\cite{Steiner10} of neutron star atmospheres concludes that the neutron star radius must be less than about 12.5 km.  Such compact stars, however,  are incompatible with the stiff 1998 RMF Shen EoS. 
Hence, the large uncertainty due the EoS indicated in Figure \ref{fig:4} is an overestimate of the EoS uncertainty. }   

Indeed, the set of equations of state that can satisfy both the neutron-star maximum mass constraint and the neutron-star radius constraint is rather limited.\cite{Abbot18}   This implies that there is probably less EoS uncertainty on the relic neutrino spectrum than that deduced in Refs.~[\refcite{Mathews14,Hidaka16,Hidaka18}].  Indeed, a new analysis of the uncertainties in the observed relic neutrino spectrum is warranted based upon these constraints.  

\begin{figure}
\includegraphics[width=4.0in]{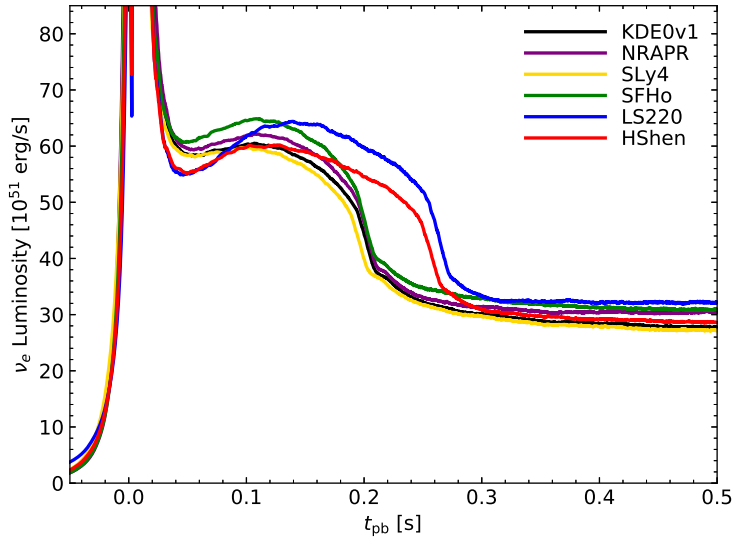}
\caption{Neutrino luminosity vs. time post bounce $t_{\rm pb}$ for various equations of state. The simulations were run using the open-source, spherically symmetric, general relativistic hydrodynamic code GR1D,\cite{GR1D} and the neutrino opacities were obtained using the open-source library NuLib.\cite{NuLib} The progenitor is the 20 M$_\odot$ star calculated by Woosley and Heger.\cite{Woosley07} The black, purple and yellow curves are the three Skyrme models from Schneider et al. 2017\cite{Schneider17} that satisfy the 2 M$_\odot$ constraint for cold neutron stars. The green curve is the SFHo EoS from Hempel et al. 2012;\cite{Hempel12} the blue curve is the EoS from Lattimer \& Swesty  1991,\cite{LS91} for an incompressibility modulus of 220 MeV; the red curve is the EoS from Shen 2011.\cite{Shen11} }
\label{fig:spec}
\end{figure}

As an illustration, Figure \ref{fig:spec} shows a new series of calculations\cite{Boccioli19}  of the emergent electron neutrino spectrum from a set of EoSs\cite{LS91,Shen11,Steiner13,Schneider17} that are mostly consistent with the neutron star maximum mass requirement.  Calculations were run with the modern fully general-relativistic spherically-symmetric open-source code GR1D.\cite{GR1D}  Neutrino luminosities were computed using the open-source library NuLib.\cite{NuLib}. The progenitor is the 20 M$_\odot$ star calculated by Woosley and Heger.\cite{Woosley07} Note that HShen refers to the  2011 version of the RMF EoS\cite{Shen11}  not the 1998 version used in Refs.~[\refcite{Mathews14,Hidaka16,Hidaka18}].  Similarly, LS220 denotes the compression modulus $K=220$ version of the Lattimer and Swesty 1991 EoS, not the $K=180$ version used previously.\cite{Mathews14,Hidaka16,Hidaka18}  Here, one can  see that there is less  uncertainty  {in the} source luminosity from the newer equations of state that satisfy the neutron star maximum mass constraint.  However,  the LS220 and HShen equations of state keep the neutrino luminosity high for longer times.  A similar conclusion has been obtained in Ref.~[\refcite{Schneider19}] for the same progenitor and GR1D CCSN model, but for 97 new Skyrme parameterized EoSs all of which satisfy observational constraints, yet show very little dispersion in the neutrino spectra.

\subsection{Dependence on Neutrino Oscillations}
It is known that the flavor eigenstates for the neutrino
 are not identical to the  mass eigenstates.  
  The flavor eigenstates $\nu_e$,
$\nu_\mu$, $\nu_\tau$ are  related to the mass eigenstates by a unitary matrix
$U$:

\begin{equation}
 \left(
\begin{array}{c}
 \nu_e \\
 \nu_\mu\\
 \nu_\tau
\end{array}
\right) 
= U
 \left(
\begin{array}{c}
 \nu_1 \\
 \nu_2 \\
 \nu_3
\end{array}
\right) ~~,
 \label{2-16}
\end{equation}
where the matrix U can be decomposed  as:
\begin{equation}
U =
\left(
\begin{array}{ccc}
  U_{e1} & U_{e2} & U_{e3}\\
  U_{\mu1} & U_{\mu2} & U_{\mu3}\\
  U_{\tau1} & U_{\tau2} & U_{\tau3}   
\end{array}
\right) \label{2-17}
\end{equation}
\begin{eqnarray}
&\equiv&
\left(
\begin{array}{ccc}
  1 & 0 & 0 \\
  0 & c_{23} & s_{23} \\
  0 & -s_{23} & c_{23}
\end{array}
\right)
\left(
\begin{array}{ccc}
 c_{13} & 0 & ~s_{13}e^{-i\delta} \\
 0 & 1 & 0 \\
 -s_{13}e^{i\delta} & 0 & c_{13} 
\end{array}
\right) \nonumber \\
&&\times
\left(
\begin{array}{ccc}
 c_{12} & s_{12} & 0\\
 -s_{12} & c_{12} & 0\\
 0 & 0 & 1 
\end{array}
\right) ~~.
 \label{2-18}
\end{eqnarray}
Here, s$_{ij}$ $\equiv$ $\sin \theta_{ij}$, c$_{ij}$ $\equiv$ $\cos \theta_{ij}$, and $\theta_{ij}$ is
the mixing angle between neutrinos with mass eigenstates i and j, and {$\delta = [0,2\pi]$ is the Dirac
 charge-parity (CP) violating  phase.} 

 For the mixing of mass eigenstates one must consider two cases:  1) the normal mass hierarchy, $m_1< m_2< m_3$;  and 2) the inverted mass hierarchy, $m_3< m_1< m_2$.  
Both models
have two resonance  density regions. The resonance at higher density is  called
the H-resonance, and the one at lower density  is called the
L-resonance. In the normal mass
hierarchy, both resonances are in the neutrino sector. In the inverted mass hierarchy, however,
one resonance is in the
neutrino sector while  the other is in the anti-neutrino sector.

 The prospects for detecting effects of neutrino flavor oscillations in the spectrum of detected SRNs has been discussed\cite{Chakraborty11} using a parameterized form for the emitted neutrino spectrum from Ref.~[\refcite{kei03b}], and also in Ref.~[\refcite{Lunardini12}] using the supernova simulations of the Basel group.\cite{Fischer10}  In Refs.~[\refcite{Mathews14,Hidaka16,Hidaka18}] the limit of 3 neutrino oscillation paradigms
were considered according to Ref.~[\refcite{Dighe00}].  

  If one  assumes an efficient conversion probability $P_H$  of $\bar\nu_3 \leftrightarrow \bar \nu_1$ at the H-resonance in the  inverted hierarchy,
then the $\bar \nu_{e}$ flux emitted from CCSNe becomes:\cite{Mathews14}
\begin{eqnarray}
\phi_{\bar \nu_e} &=&  |U_{e1}|^2 \phi_{\bar \nu_1} +  |U_{e2}|^2\phi_{\bar \nu_2} + |U_{e3}|^2 \phi_{\bar \nu_3} \nonumber \\
&=& |U_{e1}|^2 \{(1 - P_H) \phi_{\bar \nu_1}(0) + P_H \phi_{\bar \nu_3} (0)\}  \\
&+& |U_{e2}|^2 \phi_{\bar \nu_2}(0) + |U_{e3}|^2 \{P_H \phi_{\bar \nu_1}(0) + (1 - P_H) \phi_{\bar \nu_3}(0)\} \nonumber~.
\end{eqnarray}
From this, one can deduce the survival probability $\bar P$ for $\bar {\nu_e}$:
\begin{equation}
\bar P =  |U_{e1}|^2 P_H + |U_{e3}|^2 (1 - P_H) \approx 0.7 P_H ~~.
\end{equation}
The same result occurs  for the normal mass hierarchy.  Hence,  one can define this case of complete non-adiabatic mixing as: 
\begin{equation}
\bar \nu_e (0) \rightarrow 0.7 \times \bar \nu_e^0 + 0.3 \times \nu_x^0 ~~~{\rm (Case}~I).
\end{equation}

It is now known \cite{DayaBay} that the $\theta_{1 3}$ mixing is relatively large [$\sin^2{(\theta_{1 3})}= 0.092 \pm 0.0016 (stat) \pm 0.005(syst)$].  Thus, if there is no  supernova shock, then 
the survival probability can be small ($P_H \rightarrow 0$) so that the conversion of $\bar \nu_e$ into $\nu_x$ is efficient.  One can define\cite{Mathews14}  this situation of complete adiabatic mixing in the inverted neutrino mass hierarchy as:
\begin {equation}
\bar \nu_e (0) \rightarrow  \nu_x^0 ~~~{\rm (Case}~II).
\end{equation}
The case with no neutrino oscillations is adopted as Case {\it III} in Refs.~[\refcite{Mathews14,Hidaka16,Hidaka18}].

\begin{figure}[h]
\includegraphics[angle=0,width=4.5in]{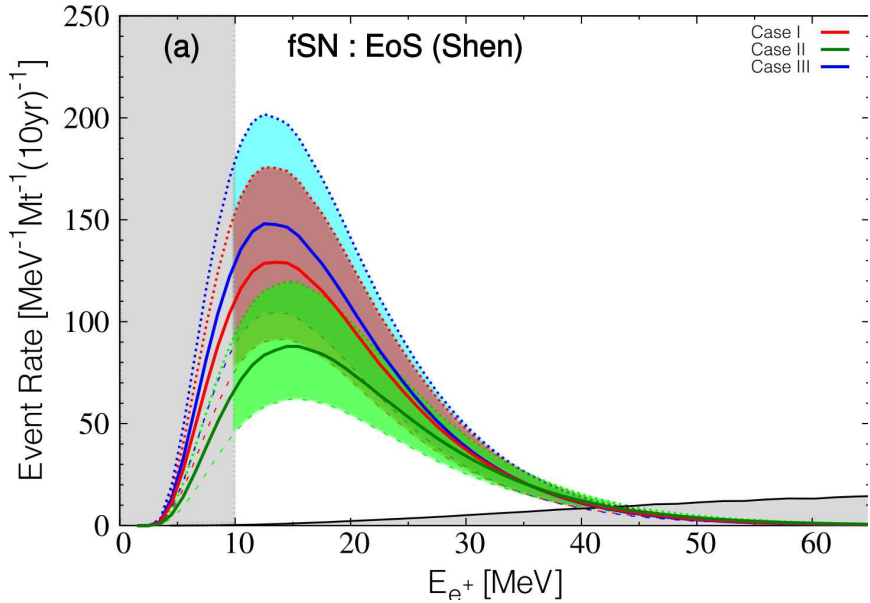}
\caption{Effects of neutrino oscillations on the predicted\cite{Mathews14} $e^+$ energy spectra  as a function of $e^+$ energy for a $10^6$ ton water \v{C}erenkov detector with 10 years of run time.  
Red, green, and blue lines shows the SRN detection rate in the case of non-adiabatic (Case {\it I}), adiabatic (Case {\it II}), or the no neutrino  oscillations (Case {\it III}), respectively for the case of a  stiff RMF\cite{Shen98} EoS.
\label{fig:6}}
\end{figure}

Figure \ref{fig:6}
illustrates the calculated\cite{Mathews14} dependence of  detection event rates on the three oscillation scenarios.  
These results  are similar to  those of other studies. 
\cite{Lunardini09,Yang11,Keehn12, Lunardini12}  Shaded bands show uncertainties from the SFR and detector statistics. 

{The calculated\cite{Mathews14} positron detection rate in Figure \ref{fig:6} shows the effect of the three different neutrino oscillation scenarios for a stiff RMF\cite{Shen98} EoS.  Here, one can see that the main effect of the neutrino oscillations is on the amplitude of the detected signal.  Whereas, the effect of the EOS dependence affects not only the amplitude but also the energy of the peak as shown in Figure \ref{fig:4}.   Thus,  it may be possible to distinguish  the oscillation  effect from the EoS uncertainty especially as the SFR  uncertainty becomes better known.}

\subsection{Dependence of SRN detection on the neutrino self interaction
  effect}
  \label{selfint}
  In the case of  an inverted mass hierarchy, a ``self interaction effect'' among  neutrinos\cite{fog07}  might cause an additional difference between the detected energy spectrum of supernova neutrinos.\cite{Lunardini12}  
The possible importance of this effect was calculated in Ref.~[\refcite{Mathews14}] for the simplest  case with a neutrino self-interaction, i.e.
a single-angle interaction. 

A single angle  interaction\cite{fog07}  causes the
 energy spectra of $\bar{\nu_e}$ and $\nu_x$ ($\bar{\nu_\mu}$
and $\bar{\nu_\tau}$) to exchange with each other at about 4.0 MeV.
This means that the energy spectra of $\bar{\nu_e}$ and $\nu_x$ are perfectly
exchanged in the detectable energy range of water \v{C}erenkov detectors.
Hence, a neutrino self interaction could change
the SRN energy spectrum in oscillation Case {\it II} into that of  Case
{\it III}. See however, Ref. \refcite{Mirizzi15} for a review of neutrino self-interactions, and the possibility that  multi angle effects suppress flavor conversions
\begin{figure}[h]
\includegraphics[width=4in]{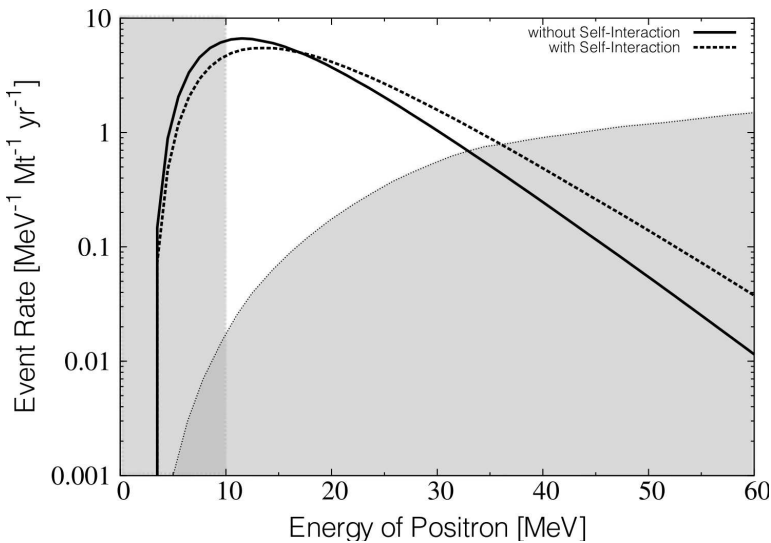}
\caption{Calculation\cite{Mathews14} of the detected SRN energy spectrum  cases with (dashed line) or without (solid line)   a single-angle neutrino self interaction.   Shaded regions are the backgrounds as defined in Fig.~\ref{fig:4}.
\label{fig:7}}
\end{figure}

Figure \ref{fig:7} from Ref.~[\refcite{Mathews14}]  shows that for the most part the correction for  neutrino self interaction produces a slight increase in the high energy end of the detected neutrino spectrum.
More recently, it has been demonstrated\cite{Ko19} that the combined effects of neutrino oscillations and neutrino self interaction can have a dramatic effect on the emergent neutrino spectrum.  Such effects may be discernible in the detected SNR spectrum and warrant further study.

\section{Conclusions}
\begin{figure}[h]
\centering
\includegraphics[width=4in]{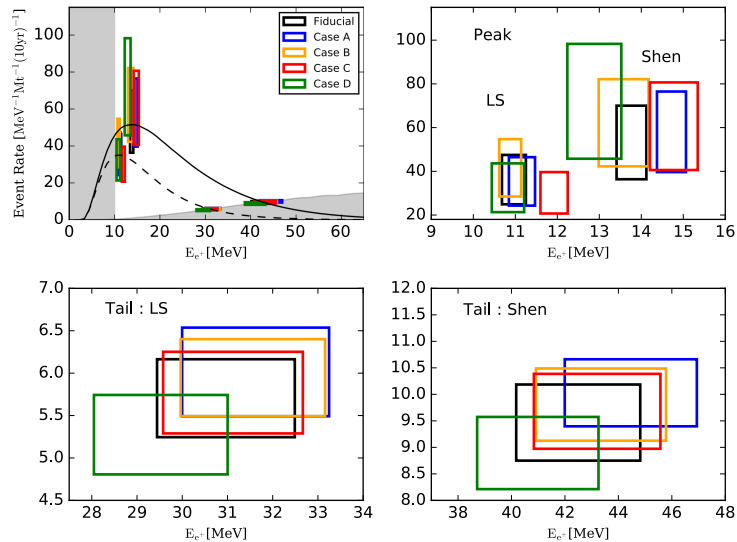}
\caption{Rectangles indicating  the location of peak and tail which characterize the spectrum for each cases. For reference,  the left upper panel shows  the spectrum of the fiducial case without the uncertainty bands included.  The solid line is for the Shen EoS and the dashed line for the LS EoS. The right upper panel shows the detail of the peak locations for both the Shen and LS EoSs. Similarly the details of tails are shown in the left bottom (for the LS EoS) and right bottom panel (for the Shen EoS).In spite of the many uncertainties, there is a unique dependence on the EoS that may be detectable in the SNR detector signal.
}\label{fig:8}
\end{figure}

\begin{figure}[h]
\centering
\includegraphics[width=4.0in]{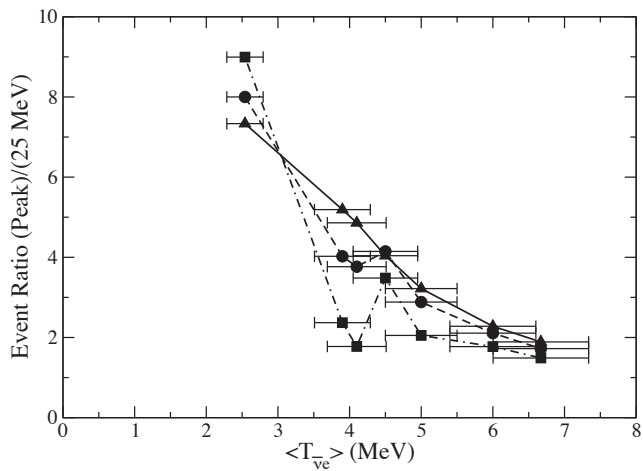}
\caption{Sensitivity or the ratio of events at the observed positron peak  to events with a positron energy of 25 MeV, corresponding to  9 fiducial SN neutrino temperatures  and different neutrino oscillation conditions adopted from the compendeum of models  in Mathews et al.\cite{Mathews14}   Circles and dashed line are for the case of neutrino oscillations with complete non-adiabatic mixing.  Squares and dot-dashed line are for the case of adiabatic mixing in the an inverted mass hierarchy.  Triangles and the solid line are for the case of no neutrino oscillations.  These points  illustrates how
the observed positron spectrum might be used to directly infer the supernova electron anti-neutrino temperature.  Error bars drawn indicate the adopted\cite{Mathews14} 10\% uncertainty in model neutrino temperatures.  The vertical scatter in the points indicates the uncertainty due to neutrino oscillation scenarios.
\label{fig:9}}

\end{figure}

The essential results of the past  studies in Refs.~[\refcite{Mathews14,Hidaka16,Hidaka18}] of the uncertainties in the predicted SRN spectrum are  summarized on Figures \ref{fig:8} and \ref{fig:9}. These works have considered a  wide variety of astrophysical  scenarios and investigated the EoS dependence of the SRN spectrum for each case with and without the occurrence of neutrino oscillations. It was consistently found that the EoS dependence of the SRN spectrum manifests prominently in the high energy tail in any scenario  considered. The robustness of this EoS dependence can be understood in that the SRN spectrum is determined mostly by the SFR for $z<2$, so that cosmological effects and metallicity-dependent effects at high redshift have a negligible effect.
However, as noted above, this EoS dependence needs to be re-evaluated in the context of the current more stringent observational constraints on the possible equations of state.

Figure \ref{fig:8} shows\cite{Hidaka18} rectangles indicating  the location of the peak and tail of the detected SRN for a variety of astrophysical conditions and the two equations of state in Figure \ref{fig:4}.    On this figure the Fiducial case is  parametrized Madau\cite{Madau14} SFR, with no neutrino oscillations;   Case A: is a revised SFR that omits lower-mass progenitors as a means to solve the red supergiant problem;\cite{Smartt09} Case B includes the  Sarburst and Quiescent SFR shown on Figure \ref{fig:4},     Case C includes a Metallicity dependent \\Variable IMF;  and  Case D includes a  SFR enhanced at high-$z$.   In that paper\cite{Hidaka18} it was concluded that the EoS sensitivity is strong enough to differentiate the EoS during CCSNe and fSNe.  However, as noted above, new constraints on the EoS may preclude this sensitivity as the soft EoS is now ruled out.

Figure \ref{fig:9} illustrates how the neutrino temperature T$_\nu$ in CCSNe explosions  can affect the detected spectra.  In particular, the neutrino temperature 
 influences the value of
the positron energy that gives the maximum
event rate, i.e. E$_{peak}$. A measurement of the ratio of the number of events at E$_{peak}$ compared to to the events at higher energy (say 25 MeV) might be used to directly infer  the neutrino temperature emerging from supernovae. The different points on this plot represent a subset of  9 points sampling the concordance region of  the compendium of 30 models published from 1987 to 2012 and listed in Table 4 of Ref.~[\refcite{Mathews14}].    These temperatures were used to infer the $\bar{\nu_e}$ spectra and associated detector signal.  We note that the most recent models suggest lower neutrino temperatures and may be associated with a steeply declining peak with energy.



\section*{Acknowledgments}
Work at the University of Notre Dame (GJM) supported by the U.S. Department of Energy under Nuclear Theory Grant DE-FG02-95-ER40934.  
This work was also supported in part by Grants-in-Aid for Scientific Research of JSPS (15H03665, 17K05459) of the Ministry of Education, Culture, Sports, Science and Technology of Japan.  



\begin{thebibliography}{0}    
\bibitem{Hartmann97}  D. H. Hartmann, D., H.  and S. E.   Woosley, Astroparti. Phys., 7, 137 (1997).
\bibitem{Totani96} T. Totani, K.Sato, and Y. Yoshida,  Astrophys. J., 460, 303 1996).
\bibitem{Beacom04} J. F. Beacom and M. R. Vagins,  Phys. Rev. Lett., 171101 (2004).
\bibitem{Beacom06} J. F.  Beacom and L. E. Strigari,  Phys. Rev. C, 73, 5807 (2006).
\bibitem{Beacom10} J. F. Beacom,  Ann. Rev. Nucl. Part. Sci. 60 439 (2010).
\bibitem{Yuksel07} H. Y\"{u}ksel, H.  and J. F.  Beacom,  Phys. Rev. D,  76, 083007 (2007).
\bibitem{Horiuchi09} S. Horiuchi, S. et al.,Phys. Rev. D, 79,  0830138 (2009).
\bibitem{Lunardini03} C. Lunardini and  A. Yu. Smirnov,  JCAP, 06, 009 (2003).
\bibitem{Lunardini09} C. Lunardini, PRL, 102, 231101( 2009).
\bibitem{Lunardini10} C. Lunardini,  arXiv, 1007, 3252 (2010).
\bibitem{Yang11}  L. Yang and C. Lunardini, Phys. Rev. D, 84, 063002 (2011).
\bibitem{Keehn12}  J. Keehn and C.   Lunardini, Phys. Rev. D, 85, 043011 (2012).
\bibitem{Lunardini12} C. Lunardiniand I.   Tamborra,  JCAP, 07, 012 (2012).
\bibitem{Chakraborty11} S. Chakraborty, S. Choubeyand K.  Kar, PLB, 702 209 (2011).
\bibitem{Nakazato13} K. Nakazato, K. PRD, 88, 083012 (2013).
\bibitem{Mathews14} G. J. Mathews,J. Hidaka, T.  Kajino,and J. Suzuki,  Astrophys. J., 790, 115 (2014).
\bibitem{Hidaka16} J. Hidaka, T. Kajino and G. J.  Mathews, Astrophys. J., 827, 85 (2016).
\bibitem{Hidaka18} J. Hidaka, T. Kajino and G. J.  Mathews,  Astrophys. J., 869, 31 (2018).
\bibitem{Nakazato15} K. Nakazato, E. Mochida, Y. Niino, and H. Suzuki, 2015, ApJ, 804, 75
\bibitem{Mirizzi15} A. Mirizzi, et al., Rivista del Nuovo Cimento Vol. 39, N. 1-2 (2016)
\bibitem{Wei16} H. Wei, Z. Wang and S. Chen, Physics Letters B 769 (2017)
\bibitem{Barranco18} J. Barranco, A. Bernal and D. Delepine, J. Phys G, 45 055201 (1918).
\bibitem{Anandagoda20} S. Anandagoda, D. Hartmann, M. Ajello and A. Desai, Res. Notes AAS, 4, 4 (2020).
\bibitem{Kyutoku17} K. Kyutoku, K. Kashiyama,  Phys. Rev. D 97, 103001 (2018).
\bibitem{Schilbach18} T. S. H. Schilbach, O. L. Caballero and G. C. McLaughlin, G. C., Phys. Rev. D 100, 043008 (2019).
\bibitem{Lin20} Z. Lin and C. Lunardini, Phys. Rev. D 101, 023016 (2020).
\bibitem{Hopkins06}  A. Hopkins and J.  Beacom, J., Astrophys. J.,  651, 142 (2006).
\bibitem{Shashank19} S. Shashank, I. Tamborra and M. Bustamante, arXiv, 1912.09115 [astri-pj.HE] (2019).
\bibitem{abe11}  K. Abe,. et al.,  (Hyper-Kamiokande working group) unpublished Letter of Intent, arXiv:1109.3262 [hep-ex] (2011).
\bibitem{Takeuchi20}Y. Takeuchi, et al. (Super-Kamiokande Collaboration), Nucl. Inst. Meth.,  952A, 161634 (2020).
\bibitem{SK12} K. Bayes,  et al. (Super-Kamiokande Collaboration) Phys. Rev. D, 85, 052007 (2012).
\bibitem{Sek13} H. Sekiya, Nucl. Phys. B, 237, 111 (2013).
\bibitem{str05} L. Strigari,  et al., JCAP, 04,  017 (2005).
\bibitem{Yuksel08} H. Y\"{u}ksel, H., et al.,  Astrophys. J.,  683, L5 (2008).
\bibitem{Horiuchi11} S. Horiuchi, J. F. Beacom,  C. S.  Kochanek,  et al., Astrophys. J., 738, 154 (2011).
\bibitem{Smartt09}  S. J. Smartt, Ann. Rev. Astron. Astrophys., 47, 63 (2009).
\bibitem{Smartt09a} S. J. Smartt,  J. J.  Eldridge, R. M.  Crockett  and J. R. Maund,  MNRAS,    395, 1409 (2009).
\bibitem{Smartt15} S. J. Smartt,  Proc. Astr. Soc. Pac.,     32, 16 (2015).
\bibitem{Sukhbold19} T. Sukhbold, T. Ertl, S. E.  Woosley,  J. M. Brown and H.-T. Janka, Astrophys. J., 821, 38 (2016).
\bibitem{Heger03} A.  Heger et al.,   ApJ 591, 288 (2003).
\bibitem{Fryer99} C. L. Fryer, S. E. Woosley and D. E.  Hartmann, ApJ, 526, 152 (1999).
\bibitem{Podsiadlowski92} P. Podsiadlowski, P. C. Joss and J. J.  Hsu, ApJ, 391, 246 (1992).
\bibitem{Nomoto95} K. Nomoto, K. Iwamoto and T.  Suzuki,  Phys. Rep., 256, 173 (1995).
\bibitem{Ugliano12} M. Ugliano, H.-T. Janka, A. Marek and A.  Arcones, Astrophys. J., 757, 69 (2012).
\bibitem{Pejcha15} O.  Pejcha and T. A.  Thompson, Astrophys. J., 801, 90 (2015).
\bibitem{Sukhbold16} T. Sukhbold, T. Ertl, S. E.  Woosley,  J. M. Brown and H.-T. Janka, Astrophys. J., 821, 38 ( 2016).
\bibitem{Couch18} S. M. Couch, M. L.  Warren and  O'Connor, E. P.,  Astrophys. J.,  890, 127 (2020).
\bibitem{Isern91} J. Isern, R. Canal and L.  Labay, Astrophys. J.l, 372, L83 (1991).
\bibitem{Madau14} P. Madau and M.  Dickinson,  ARA\&A, 52, 415 (2014).
\bibitem{Lacey16} C. G. Lacey, C. M.  Baugh, C. S.  Frenk, C. S., et al.  MNRAS, 462, 3854 (2016)
\bibitem{Conroy13} C. Conroy and P. G.  van Dokkum, ApJ, 760, 71 (2012).
\bibitem{RR_SFR_2016} M. Rowan-Robinson, S.  Oliver, L. Wang, et al., MNRAS, 461, 1100 (2016,).
\bibitem{kob00}  C. Kobayashi, C., et al., Astrophys. J.  539, 26 (2000).
\bibitem{Baldry03}  I. K. Baldry and  Glazebrook, ApJ, 593, 258 (2003).
\bibitem{Padoan02} P. Padoan and A. Nordlund,  Astrophys. J., 576, 870 (2002).
\bibitem{Martin-Navarro_2016} I. Mart\'in-Navarro, J. P.  Brodie, R. C. E.  van den Bosch, A. J. Romanowsky and D. A.  Forbes, Astrophys. J., 832, L11 (2016).
\bibitem{Ma_2016} X. Ma, P. F. ,Hopkins, C.-A.,Faucher-Gigu\'ere, et al.,  MNRAS, 456, 2140 (2016).
\bibitem{Lopes_2014} A. R. Lopes,  A. Iribarrem,  M. B. Ribeiro and W. R. Stoeger,  A\&A, 572, A27 (2014).
\bibitem{Lattimer12} J. M. Lattimer,  Ann. Rev. Nucl. Part. Sci, 62, 485 ( 2012).
\bibitem{arn89} D. Arnett,  et al.,  ARA\&A, 27, 629 (1989).
\bibitem{OConnor18} E. O'Connor et al., J. Phys. G, 45, 104001 (2018).
\bibitem{Mayle87} R. Mayle,  et al.,  Astrophys. J., 318, 288 (1987).
\bibitem{Wilson03} J. R. Wilson and G. J. Mathews 2003, J. R. Wilson and G. J. Mathews, {\it  Relativistic Numerical Hydrodynamics}, (Cambridge University Press,2003).
\bibitem{Olson16} J. P. Olson, et al., 2016 [arXiv:1612.08992].
\bibitem{str03b} A. Strumia,  and F.   Vissani,  PLB, 564, 42 (2003).
\bibitem{LS91}  J. Lattimer and F. D.  Swesty,   Nucl. Phys. A, 535, 331 (1991).
\bibitem{Shen98}  H. Shen, H., et al., Nucl. Phys. A, 637, 435 (1998)
\bibitem{Sumiyoshi05} K. Sumiyoshi, K., et al., Astrophys. J., 629,  922  (2005).
\bibitem{Sumiyoshi08} K. Sumiyoshi, K., et al.,  Astrophys. J.,  688, 1176 (2008).
\bibitem{Demorest10} P. B. Demorest, T. Pennucci, S. M. Ransom, M. S. E. Roberts, and J. W. T.  Hessels,  Nature. 467, 1081-1083 (2010).
\bibitem{Antoniadis13} J. Antoniadis,  et al. Science. 340, 448 (2013).
\bibitem{Cromartie19} H. T. Cromartie, et al., arXiv: 1904.06759 (2019).
\bibitem{Abbot17}B. P. Abbot, et al. {\it LIGO and Virgo Scientific Collaboration} PRL, 119, 161101 (2017).
\bibitem{Abbot18}B. P. Abbot, et al. {\it LIGO and Virgo Scientific  Collaboration} PRL, 121, 161101 (2018).
\bibitem{Steiner10} A. W. Steiner, J. M. Lattimer, and E. F. Brown, Ap.J. 722, 33 (2010).
\bibitem{Boccioli19} L. Boccioli, G. J. Mathews, and E. O'Connor, ApJ Submitted (2019). 
\bibitem{Shen11} H. Shen, H. Toki, K. Oyamatsu, and K. Sumiyoshi, ApJ, 197, 20 (2011).
\bibitem{Steiner13} A. W. Steiner, JM. Hempel and T. Fischer, ApJ,  774, 17  (2013).
\bibitem{Schneider17} A. S. Schneider, L. F. Roberts, and C. D. Ott,  Phys. Rev. C 96, 065802 (2017).
\bibitem{GR1D} E. O'Connor and C. D. Ott, Class. Quant. Grav., 27, 114103 (2010).
\bibitem{Boccioli19} L. Boccioli, G. J. Mathews, M. Warren, I. Suh and M. Correa,  Bull. Am Phys. Soc.,  April, T08.008 (2019).
\bibitem{NuLib} E. O'Connor,  Astrophys. J. Suppl. Ser., 219, 24 (2015).
\bibitem{Itoh75} N. Itoh, Progr. Theor. Phys.,  54, 1580 (1975).
\bibitem{Bruen85} S. W. Bruenn, Astrophys. J. Suppl. Ser., 58 771 (1985).
\bibitem{Horowitz02} C. J. Horowitz, Phys. Rev. D 65 043001 (2002)
\bibitem{Horowitz97} C. J. Horowitz, Phys. Rev. D 55 457781 (1997)
\bibitem{Bruenn97} S. W. Bruenn  and A. Mezzacappa,  Phys. Rev. D 56 7529 (1997).
\bibitem{Rampp02} M. Rampp and H.-T. Janka, Astron. Astrophys. 396 361 (2002).
\bibitem{Hannestad98}  S. Hannestad and G. Raffelt, Astrophys. J. 507 339 (1998).
\bibitem{Woosley07} S. E., and A. Heger,  Phys. Rep., 442, 269 (2007).
\bibitem{Schneider19} A. S. Schneider, L. F. Roberts, and C. D. Ott, and E. O'Connor,  arXiv:1906.02009 (2019).
\bibitem{Hempel12} M. Hempel, T. Fischer, J.  Schaffner-Bielich and M.  Liebend\"orfer, Astrophys. J. 748, 70 (2012).
\bibitem{kei03b}  M. T. Keil, G. G.  Raffelt and H.-T. Janka, , Astrophys. J., 590, 971 (2003).
\bibitem{Fischer10} T. Fischer,  et al.,  A\&A, 517 A80 (2010).
\bibitem{Dighe00} A. S. Dighe and A. Y.  Smirnov,  Phys. Rev. D, 62, 3007 (2000).
\bibitem{DayaBay} F. P. An, et al. (Daya Bay Collaboration), Phys. Rev. Lett., 108, 171803 (2012).
\bibitem{fog07}  G. Fogli, et al., JCAP, 12, 010 (2007).
\bibitem{Ko19} H. Ko, et al.,  Astrophys. J. Lett., 891, L24 (2020).


\end{thebibliography}
\end{document}